# Techniques frequently used during London Olympic judo tournaments: A biomechanical approach

S. Sterkowicz,[1] A. Sacripanti[2], K. Sterkowicz – Przybycien[3]

1  Department of Theory of Sport and Kinesiology, Institute of Sport, University School of Physical Education, Kraków, Poland
2  Chair of Biomechanics of Sports, FIJLKAM, ENEA, University of Rome "Tor Vergata", Italy
3  Department of Gymnastics, Institute of Sport, University School of Physical Education, Kraków, Poland

**Abstract**
Feedback between training and competition should be considered in athletic training. The aim of the study was contemporary coaching tendencies in women's and men's judo with particular focus on a biomechanical classification of throws and grappling actions.
359 throws and 77 grappling techniques scored by male and female athletes in Olympic Judo Tournaments (London 2012) have been analyzed. Independence of traits (gender and weight category by technique classes) was verified via $c^2$ test. Comparison between frequency of each subsequent technique class and rest/inconclusive counts was made in 2×2 contingency tables. The significance level was set at p£0.05.
Throwing technique frequencies grouped in the seven biomechanical classes were dependent on gender. A significant difference was found between frequencies of variable arm of physical lever technique scored by males (27.09%) and females (16.67%) as compared to the rest/inconclusively techniques counts. Significant differences between men who competed in extra lightweight and heavy weight concerned the frequency of the techniques used with maximum arm or variable arm of physical lever and a couple of forces applied by trunk and legs. In females, a tendency to higher frequency of techniques that used couple of forces applied by arm or arms and leg was observed
in extra lightweight compared to the heavy weight.
Because the technique preferred in the fight depends on a gender and weight category of a judoka, the relation- ships found in this study, which can be justified by the biomechanics of throws, should be taken into consideration in technical and tactical coaching of the contestants. A method used in this study can be recommended for future research concerning coaching tendencies.

## Background

Observation of the fight in judo competitions is necessary as it is the only opportunity for verification of the process of the contestant's coaching. This feedback is particularly important if new rivals emerge in a weight category or if fighting regulations force athletes and coaches to face new, more demanding challenges [1].
World championships and the Olympic Games are the moments that best summarize many years of training. Male [2,3] and, more recently, female [4] competitions have been analysed in detail during these events, both in groups and individually [4]. These analyses used a traditional classification of techniques developed in the Kodokan Judo Institute [5], with throws including hand techniques, loin techniques, foot and leg techniques and the art of throwing in a horizontal posture, sacrificing his own body's balance. In addition, the art of grappling encompasses holds, strangling, and locking the joints (elbows). The need for improvement of technical and tactical preparation of contestants has caused that some additional criteria of joint classification have also been used in practice: based on the direction of kuzushi (the action taken by one contestant in order to throw the other contestant out of balance), presence of body rotation performed by a thrower and tactical situation when performing a throw i.e. a single attack, combination or counterattack [6]. The observations and analysis of the course of the fight during competitions at the elite level have been the focus of studies that yielded results that are useful for both theory and practice of judo and judo coaching. At the same time, a consistent throw classification was developed based on the biomechanical criteria [7,8], which has not been used for the analysis of the frequency of actions in a judo fight so far.

## Concept of the work

All judo athletes used the same fighting rules regardless of whether they were males or females. It was assumed that the frequency found for a particular technique class might be related to gender or a weight category. Therefore, it seemed justified to formulate the following research questions: (1) Which throw techniques and grappling actions (Sacripanti's biomechanical criterion) are frequently used by contemporary elite judo athletes?; (2) Were there any differences between the frequency observed in male and female competitors?; (3) Were there any differences depending on weight categories?; (4) How often did the penalties for non-combativity in judo fights occur? The aim of the present study was contemporary coaching tendencies in women's and men's judo with particular focus on biomechanical classification of throws.

## Material and Methods

There are no ethical issues involved in the analysis and interpretation of the data used as these were obtained from other sources and were not generated by experimentation. The athletes' personal identification was re- placed by a code, which ensured anonymity and confidentiality. All actions of male and female athletes were recorded using IJF coding system [9]. There were 359 judo-throw-techniques and 77 grappling actions per- formed and scored during Olympic tournaments in London. Penalties caused by the breach of judo fight regulations (n=591) were also analyzed. Furthermore, each technique was rearranged by us into the biomechanical classification system [7,8,10]. A data analysis was conducted for identification of each technique within nine classification groups (Tables 4A and 4B in Annex [8,11,12]). Frequency of technique count distribution was compared using Statgraphics Centurion XVI.I software. Independent variables were gender (males; females) and weight categories: (1) extra light- weight; (2) half lightweight, lightweight, half middle-weight; (3) middleweight; half heavyweight; (4) heavy weight. The rationale behind this division was separation of semiopen categories i.e. 1st group (upper limits, only) and 4th group (lower limits, only). The ranges between the limits in the 2nd group were 21.0% and 19.0% of the lower limit of heavy weight of men and female. In the 3rd group, these ranges were similar (19.0% and 19.2%, respectively). In the multi-way tables, due to the expectedly small numbers, independence of traits was verified with $c^2$ test in the logarithmic form (G-test) [13]. Comparisons between the frequency of each subsequent technique classes and a rest counts were presented in 2×2 contingency tables. The Yates correction verified by Fisher exact test (FET) was used for small data. In the case of the significant dependency, the contingency coefficient C was calculated. The significance level was set at p£0.05.

## Results

Table 1 presents the frequency of techniques used by males and females from different weight categories. In general, techniques based on a couple of forces were used less frequently (39.6%) than the techniques used with physical lever (60.5%). There were no significant differences (c2=0.875, df=1, p=0.350) between the frequencies of these techniques performed by males (37.4% vs. 62.6%) and females (42.3% vs. 57.7%). Technique frequencies grouped in the seven biomechanical classes (Table 1) were dependent on gender (c2=16.00, df=6, p<0.05, C=0.207).

The techniques used with maximum physical lever were scored the most often (25.1%), independently of the competitor's gender (p>0.05), i.e. 24.1% in male and 26.3% in female judokas. In those techniques, the group of tai-otoshi was a typical throw scored (7.0%) with similar frequencies in males and females (6.4% and 7.7%, respectively). Using the biomechanical criterion, the next frequently scored throws were those using a variable arm of physical lever (22.6%). Significant differences (c2=5.24, df=1, p=0.022, C=0.120) were found between the frequencies of this class of techniques scored by males (27.1%) and females (16.7%) as compared to the rest/inconclusively performed techniques count. Seoi-nage is an example of throw (14.8%) which was less frequently used by females (8.3%) compared to males (19.7%). The frequencies of other techniques classified within next five classification groups (see Table 1)

**Table 1.** Frequency of throw techniques used during Judo Olympic Tournaments (2012) by males and females from different weight categories.

| Technique of throws codes CODE | Total | Males | Females | Male groups/weight categories | | | | Female groups/weight categories | | | |
|---|---|---|---|---|---|---|---|---|---|---|---|
| | | | | Group 1 | Group 2 | Group 3 | Group 4 | Group 1 | Group 2 | Group 3 | Group 4 |
| PLmaxA | 90 | 49 | 41 | 9 | 24 | 14 | 2# | 8 | 14 | 12 | 7 |
| PLvA | 81 | 55 | 26* | 17 | 23** | 14** | 1** | 2 | 13 | 8 | 3 |
| CAL | 77 | 40 | 37 | 9 | 17 | 11 | 3 | 6 | 15 | 15## | 1# |
| CTL | 55 | 27 | 28 | 2 | 12 | 8 | 5* | 1 | 12 | 12 | 3 |
| PLminA | 35 | 20 | 15 | 2 | 7 | 8 | 3 | 1 | 7 | 3 | 4 |
| PLmidA | 11 | 3 | 8 | 0 | 1 | 1 | 1 | 0 | 4 | 2 | 2 |
| CA | 10 | 9 | 1 | 0 | 6 | 2 | 1 | 0 | 0 | 0 | 1 |
| Total | 359 | 203 | 156 | 39 | 90 | 58 | 16 | 18 | 65 | 52 | 21 |

*PLmaxA – Physical lever applied with max arm; PLvA – Physical lever applied with variable arm; CAL – Couple of forces applied by arm or arms and leg; CTL – Couple of forces applied by trunk and legs; PLminA – Physical lever applied with min arm; PLmidA – Physical lever applied with middle arm ; CA – Couple of forces applied by arms. * Significant difference between males and females; # significant difference between group 1 and group 4; ** significant difference between group 1 and 2, 3, 4 groups, ## significant difference between groups $2^{nd}$ and $3^{rd}$*

A comparison between a particular technique and the rest/inconclusive techniques in females shows a tendency to higher frequency of CAL technique used in 1st compared to the 4th group ($c2=3.608$, $df=1$, $p=0.057$, $C=0.291$; $FET=0.035$). The CAL technique was also relatively often used in 3rd compared to the 4th weight category ($c2=3.760$, $df=1$, $p=0.053$, $C=0.176$; $FET=0.029$). In gripping actions, women lost because of the vascular chokes more often than men (Table 2). Men from the 1st weight category performed pin-ning techniques of the four corner type ($c2=4.024$, $df=1$, $p=0.045$, $C=0.191$) much less frequently than those from the 3rd category. Among women, the number of particular grappling techniques did not depend on weight categories.

The frequency of penalties for non combativity was significantly higher among men (65.4%) than in women (55.5%, cell's percentage of the column) ($c2=9.783$, $df=1$, $p=0.002$, $C=0.128$). The men from extra light-weight category were imposed penalties for non-combativity less frequently (56.8%) than those from the heavy weight (83.9%) ($c2=8.797$, $df=1$, $p=0.003$, $C=0.286$). Similar results were observed for comparison of the 2nd (58.3%) and 4th groups (83.9%) ($c2=12.985$, $df=1$, $p<0.001$, $C=0.236$) as well as the 3rd (68.6%) and 4th groups (83.9%) ($c2=4.703$, $df=1$, $p<0.030$, $C=0.167$). No relationships were found between the frequency of penalties and weight category in women (Table 2).

Table 2. Frequency of grappling techniques used during Judo Olympic Tournaments (2012) by males and females from different weight categories.

| Grappling actions codes | Total | Males | Females | Male groups/weight categories | | | | Female groups/ categories weight | | | |
|---|---|---|---|---|---|---|---|---|---|---|---|
| CODE | | | | Group 1 | Group 2 | Group 3 | Group 4 | Group 1 | Group 2 | Group 3 | Group 4 |
| KGARAMI | 1 | 0 | 1 | 0 | 0 | 0 | 0 | 0 | 0 | 0 | 1 |
| KHISHIGI | 19 | 9 | 10 | 5 | 3 | 1 | 0 | 0 | 6 | 3 | 1 |
| OKESA | 11 | 7 | 4 | 2 | 2 | 1 | 2 | 0 | 3 | 1 | 0 |
| OSHIHO | 30 | 19 | 11 | 2 | 9 | 6# | 2 | 1 | 2 | 5 | 3 |
| SRESP | 12 | 8 | 4 | 2 | 4 | 0 | 2 | 1 | 3 | 0 | 0 |
| SVASC | 4 | 0 | 4* | 0 | 0 | 0 | 0 | 2 | 2 | 0 | 0 |
| Total | 77 | 43 | 34 | 11 | 18 | 8 | 6 | 4 | 16 | 9 | 5 |

KGARAMI – joint techniques of the entangled joint lock type; KHISHIGI – joint techniques of the bending and pressing against elbow joint type; OKESA- pinning techniques of the scarf type; OSHIHO – pinning techniques of the four corner hold; SRESP – respiratory chocking; SVASC – vascular chocking. * Significant difference between males and females, # significant difference between group 1st and group 3rd.

**Table 3.** Number of penalties imposed during fights of men and women according to weight categories.

| Instances of penalties and codes CODE | Total | Males | Females | Male groups/weight categories | | | | Female groups/weight categories | | | |
|---|---|---|---|---|---|---|---|---|---|---|---|
| | | | | Group 1 | Group 2 | Group 3 | Group 4 | Group 1 | Group 2 | Group 3 | Group 4 |
| P29 Non-combativity | 364 | 238 | 126* | 21 | 95 | 70 #, ## | 52 ** | 20 | 59 | 31 | 16 |
| Other penalties | 227 | 126 | 101 | 16 | 68 | 32 | 10 | 15 | 55 | 17 | 14 |
| Total | 591 | 364 | 227 | 37 | 163 | 102 | 62 | 35 | 114 | 48 | 30 |

* Significant difference males from females, # Significant difference between group 1st and group 4th, ** significant difference between group 2nd and group 4th, ## significant difference between group 3rd and group 4th.

## Discussion

*Hierarchy of throw techniques scored by male and female judo Olympians*

Judo athletes preferred PLmaxA techniques. They per- formed these techniques more often than PLmidA, particularly PLminA techniques. From the biomechanical point of view, the force with the same magnitude and direction that acts on the greater lever causes greater effect (moment of force). The frequency of performing the above techniques depended neither on gender nor on weight category. This status reflects the principle that is used in technical and tactical preparation of judokas i.e. „*Maximum-Efficiency with Minimum Effort*" [5]. The underlying idea of judo declares the possibility of winning with opponents with greater physical strength. According to this principle, technical excellence means using the strength and inertia of the opponent against them [2]. In general, at equal resistance, when the arm of the lever used in a lever technique increases the applied force decreases. This means that lever techniques of maximum arm are energetically most effective among lever techniques group. But more subtle information can be derived from this analysis on fighting rhythm. In general, throwing techniques are connected with shifting velocity of Athletes couple system during the fight. In fact, using whatever lever techniques tori (attacker) needs for a while to stop himself to properly apply the technique. In the last Olympic Games, the rhythm of a fight was relatively quiet, also caused by the high non-combativity (65.4% for males and 55.5% for females). Coordinative and strength athletes' capabilities were also increased thanks to the increasingly advanced training methodologies. This happens because the lever techniques are more complex (Figure 1, in Annex) as the whole movement, and they need higher coordination of the body and kinetic chains [GAI + (SSAI + ISAI) + Lever + Kake], for example seoi-nage (Figure 2) compared to couple techniques or [GAI + Couple + Kake] uchi-mata (Figure 3), but they are also more energy-consuming, as already demonstrated in many specific papers [10,14–17].

Men used the PLvA technique more often than women. The lower frequency of the effective PLvA techniques used by women was probably caused by lower upper body strength reflected in bench press and rowing tests [18]. Another group of judokas was characterized by higher percentage of relative torque in knee extensors, with lower percentage of flexors and trunk extensors compared to untrained controls. Although judo contestants exhibit similar relative strength to untrained peers, many years of training cause that they demonstrate higher strength in the muscles that are active when pulling or lifting the opponent when performing throws. Antigravity muscles are able to develop particularly high force in these people: they play an essential role when throws are performed [19]. Individual body build characteristics and experience cause that strength profile in elite seniors was connected with the preferred techniques of performing throws (foot and leg techniques or hand techniques) [20]. An explanation of the difference observed in the frequency of physical variable arm level techniques (PLvA) between males and females can be provided with an example of a seoi-nage throw. When shoulder throws such as seoi-nage are performed, a compensation of body posture can be observed, connected with disproportions in the status of force development. With knee extensors weaker than hip extensors, smaller knee bend is naturally observed. Lifting opponents will occur with unfavorable position of inclination forward (longer lever arm for the acting force). This situation is typical of weak antigravity muscles in lower extremities, both knee and hip extensors. If an athlete's knee extensors are weaker than those in hips, this state can be compensated by higher hip bend angle, without the necessity to incline the body trunk [19]. The relatively high flexibility in female kinetic chains often compensate for the weaker knee extensors with the helping application of makikomi supplementary movement in PLvA application, however the weaker arm strength and, in general, different hip and gluteus dimensions make it very difficult to use these techniques fast and explosively.

***Differences in throw techniques scored between extra lightweight and heavy weight male athletes***

Significant differences between men from extra light-weight and heavy weight categories were found in PLmaxA, PLvA and CTL techniques. In addition, an increase in contingency coefficient strength was ob-served between 1st and next consecutive weight categories, i.e. 2nd, 3rd and 4th. The proportionality of stature that changes with weight category is likely to have a particular importance. Heavyweights are usually proportionally shorter than lightweights, i.e. they are less ectomorphic than lightweights [21]. The body proportionality of an athlete should be related to his/her techniques preferred [22]. It is essential for judo that a compromise between keeping optimal body weight and composition and both physiological and motor efficiency is obtained [23]. Many results obtained for fat percentage in judo-kas were evaluated using different equations. However, the research carried out by the same authors and using the same methods [24] demonstrated increased adiposity in judokas from heavier weight categories. Those results corroborated findings of Callister et al. [25], who found moderate correlations between body mass and percent fat.

Relative dimensions of trunk and the differences in body mass and relative arms' strength are related to the significant differences between men from extra lightweight and heavy weight categories application of PLmaxA, PLvA and CTL techniques. In general, the heavyweights like to apply couple techniques that are simpler and energy-efficient.

As mentioned above, lever techniques are more complex as the whole movement, but they also need higher coordination of the body and kinetic chains [GAI + (SSAI + ISAI) + Lever + Kake] compared to the couple techniques. [GAI + Couple + Kake]. In terms of the fight, this means that heavyweights, who prefer qui-et pace during a contest, apply these relatively simpler techniques with high velocity to shorten the distance and fast application of couple. The bigger trunk dimensions support the CTL use because it promotes essential mechanics of this kind of techniques (to move the heavy adversary's body around his center of mass). On the contrary, they obviously have more difficulty in ap-plying both PLmaxA and PLvA than competitors from extra lightweight categories, because, in general, the coordinative capabilities are lesser than in the lighter categories, but also because the essential mechanics for the physical lever techniques is the result of a well-coordinated and well-interconnected action performed by both kinetic chains in different time sequences that aims to translate the adversary's centre of mass in space [10].

First, there is a superior-chain open space that involves the body as part of the opponent's grip; secondly, there is the general action (reducing the distance) that is pursued and harmonically followed up by the coordinated and connected work of both Inferior and Superior Action Invariants as achieved through the abdominal and trunk muscles. These techniques need more skill in harmonic chains-connected movements, than Couple techniques; in fact, such techniques are often ineffective because of a lack in harmony in one of the preceding movements halts the throw, essentially preventing any score. Obviously such harmonic-complex movements are easier for extra light weights than for heavy ones.

The body composition factor can interact with the preference for a particular technique performed by heavy- weights or lightweights, as relative strength tends to be frequently lower in heavyweights than in lightweights. High resistance in training and competition during many years of sport-selection process is likely to cause changes in body build. A very low difference of sexual dimorphism index was observed for height-weight ratio, ecto-morphy, fat free mass percentage and calf girth. Average index in untrained subjects was higher than in judokas [26]. More often for female athletes, application of throws is Innovative or Classic, very few Chaotic Forms [10] are seen in women competition, but the percent- age of Innovative variations is higher due to their body's flexibility. Connection tachi-waza , ne-waza, for koshi- waza is very often linked to the application of makikomi variation of throwing techniques. Normally, in women competitions, grip fight is less strength-based, while the attack velocity is not as explosive as in men competition. It is interesting to note that the poor presence of Chaotic Form of techniques in women games is directly connected to the natural and relative lack of strength both in hands and legs of female athlete's body structure. Therefore, women's judo remains more connected to classic Kodokan Judo as for grips preference and the form of throwing techniques applied (Classic or Innovative) [10].

Relatively more vascular chokes instances were observed in women compared to men. This is likely to be connected to the unified training methods that aim to in- crease arms strength both in male and female athletes for grip goals. It is common knowledge that the muscular force generated in arms by women could increase with strength training to the level of 85–90% of the values recorded in men with similar weight, although this in- formation has not been validated by any scientific studies. Recent studies, however, [27] found that female elite athletes (involving well-trained judo athletes) had lower hand grip strength than the untrained male subjects.

Furthermore, the change in arm strength is very often not connected to the similar increase in other muscle groups like neck, which in female athletes is probably weaker, with less developed sternocleido-mastoid muscles that protect carotid arteries in women and man. This problem needs a biomechanical research in a future.

Unexpectedly, the frequency of penalties imposed for non- combativity was significantly higher in the group of men than in women. It can be associated with the differences in psychological preparation rather than physical one [28,29].

## Effectiveness

Interesting also is the evaluation of effectiveness of throwing techniques applied in the London Olympic Tournament.

Both for male and female the preference is for a bit for the couple group than for the lever techniques, more effective the female both into the couple and lever application this very subtle difference could be justified on the less defensive capability that is present in female group than in the more effective application of the techniques. ( see tab 4)

| Throws Effectiveness In London Olympic 2012 | | |
|---|---|---|
| *Throws* | *Effectiveness Male %* | *Effectiveness Female %* |
| Seoi ( Ippon – Morote - Eri) | 14.8 (329) | 8.2 (222) |
| Uchi Mata | 9.2 (138) | 15 (143) |
| O Uchi Gari | 15 (53) | 24 (49) |
| Ko Uchi Gari | 12 (57) | 37 (35) |
| Tai Otoshi | 25 (36) | 23.8 (21) |
| Soto Makikomi | 10 (10) | 23.6 (17) |
| Tani Otoshi | 46 (13) | 50 (16) |
| Uchi Mata sukashi | 90 (10) | 100 (10) |
| *Couple* | *28.7* | *39* |
| *Lever* | *24* | *26.4* |

*Tab.4 Throws Effectiveness In London Olympic 2012*

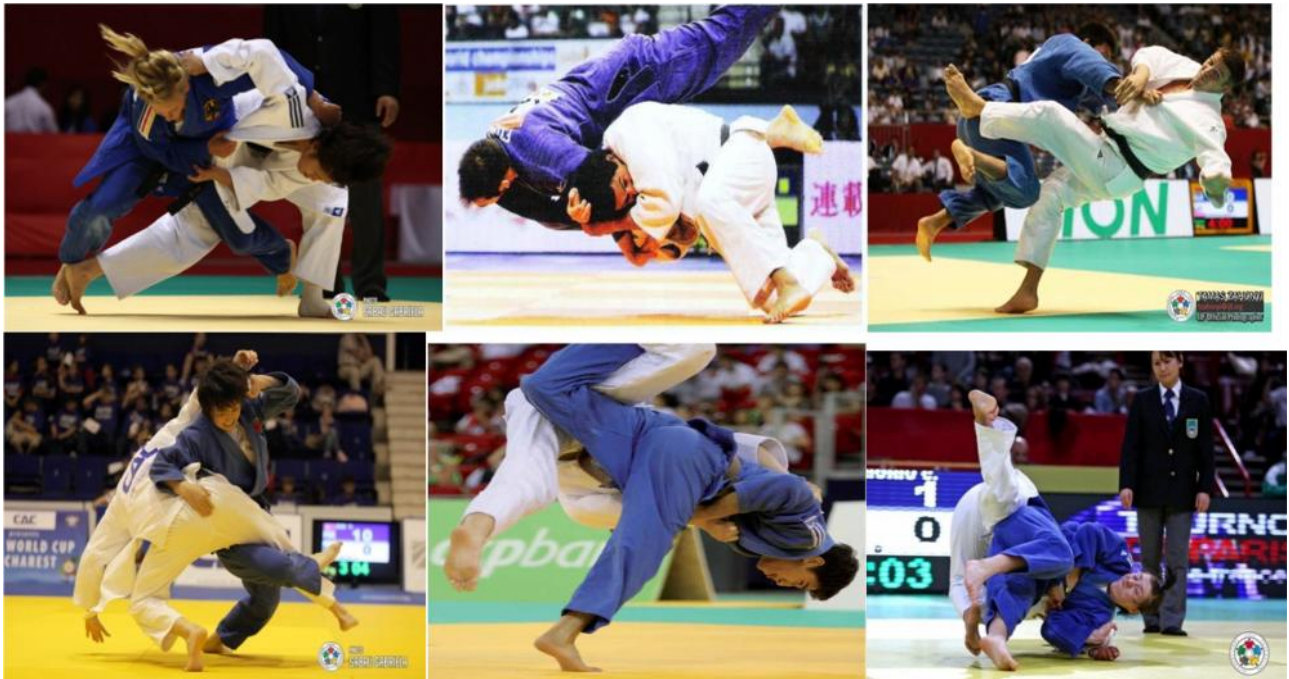

*Figg-1-6 The six more effective Innovative variations of techniques in London Olympic
: Tai Otoshi, Seoi Nage, O Uchi Gari, Ko Uchi Gari, Uchi Mata, Soto Makikomi.*

*All pictures archive IJF ( Tamas Zhaoni, Sabau Gabriela) Thanks to the IJF President Marius L.Vizer*

**Conclusions**
Because the techniques preferred during fighting depends on a gender and weight category of judokas, the relation- ships found in this study, which can be justified by the biomechanics of throws, should be taken into consideration in technical and tactical coaching of the contestants. A method used in this study can be recommended for future research concerning coaching tendencies.

## Practical applications

Statistical relationships concerning the choice of the fighting technique depending on gender and weight category were justified with biomechanics of the throws performed. Normally, it is well know that couple techniques are energetically more convenient compared to lever techniques. Body build should be considered when choosing the fighting technique, particularly when the opponent is higher or shorter or they use an opposite left or right grip kenka-yotsu. The quality of actions per- formed by the contestants should be monitored and analyzed during competitions and training in order to optimally select the means of physical preparation and
stimulate technical and tactical preparation in terms of counterattack techniques or combined techniques. In order to achieve this, it is essential to focus on individual training of a particular contestant.

Throughout the years, female judo (which was biomechanically more Kodokan classic) have approached the men's style. The observation of the London Olympic Games indicates a very unified approach to the training methodologies among male and female athletes in the world, that highlight the equivalent PLmaxA and CAL percentage in a fight, in spite of the natural differences in arm strength between male and female athletes.

# *Annexes*

**GAI** *"General Action Invariants"* – all the trajectories applied to shorten the distance between athletes [10].

**(SAI) – *"Specific Action Invariants"***, which can be split into Superior Specific Action Invariants (SSAI) and Inferior Specific Action Invariants (ISAI) all the movements performed by Athlete's kinetic chains [10]

*"Innovative Throws"* – are all throwing techniques that keep alive the formal aspect of classic judo throws, and differ in terms of grips and final direction of applied forces only [12].

"*New or Chaotic Throws*" – principally belong to the lever-type Group, and are characterized by the application of different GAI trajectory, grips positions (SSAI) which apply force in different (non- traditional) directions while simultaneously applying (ISAI) stopping point in non- classical positions [10].

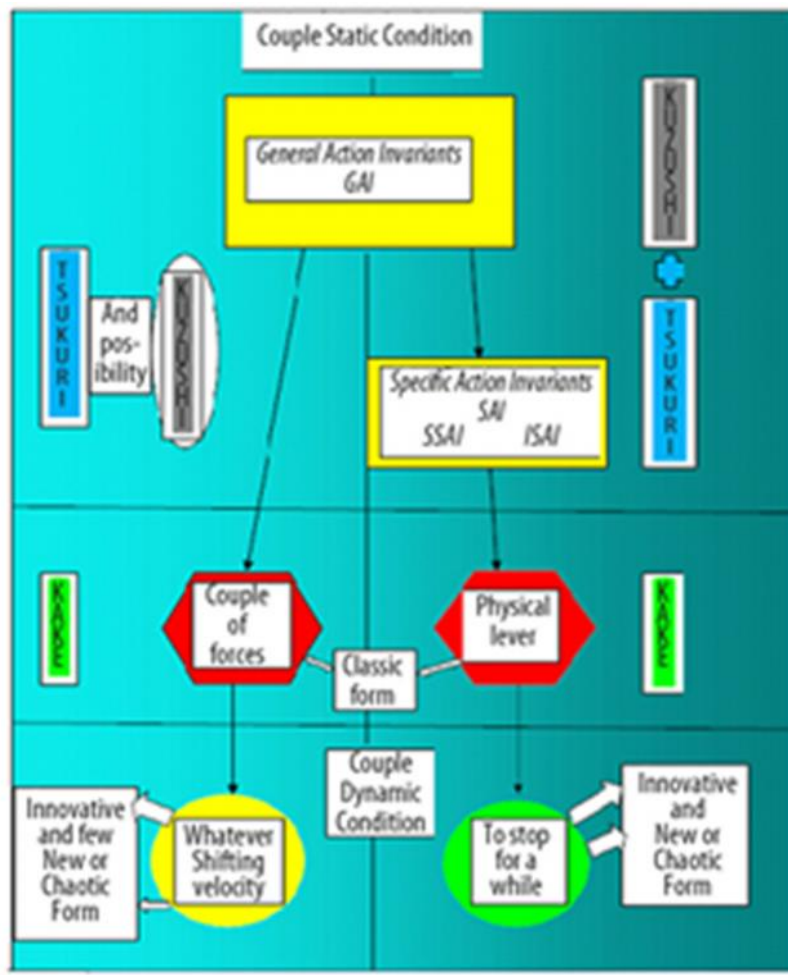

*Figure 1 annexes. Summary of the Kuzushi Tsukuri Action Invariants connected to Kake phase and Classic or Innovative and New (or Chaotic) Form of throwing techniques [10]. The figure is based on the most recent Kodokan classification [11] and innovative techniques names [12] and others*

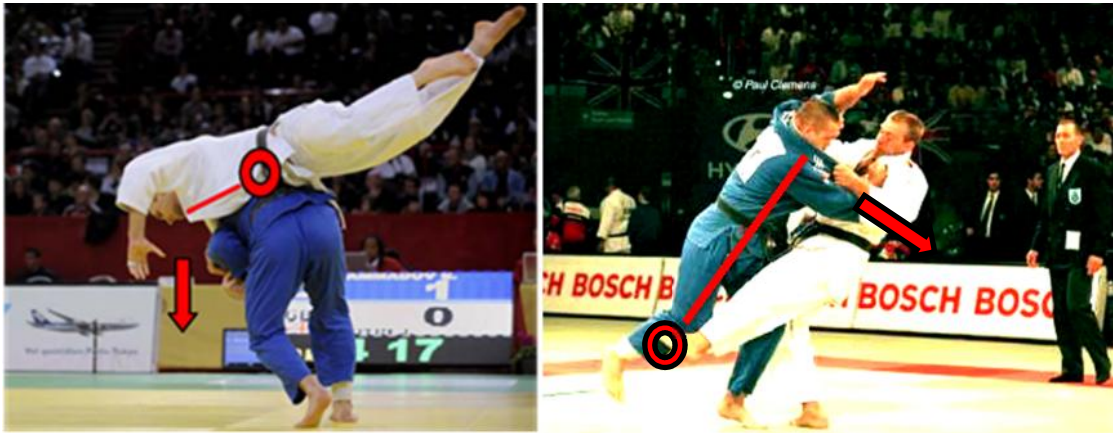

*Fig 2,3 annexes Seoi-nage is typical technique of Physical lever applied with variable arm [8]. Hiza guruma lever with medium arm (David Finch with permission).*

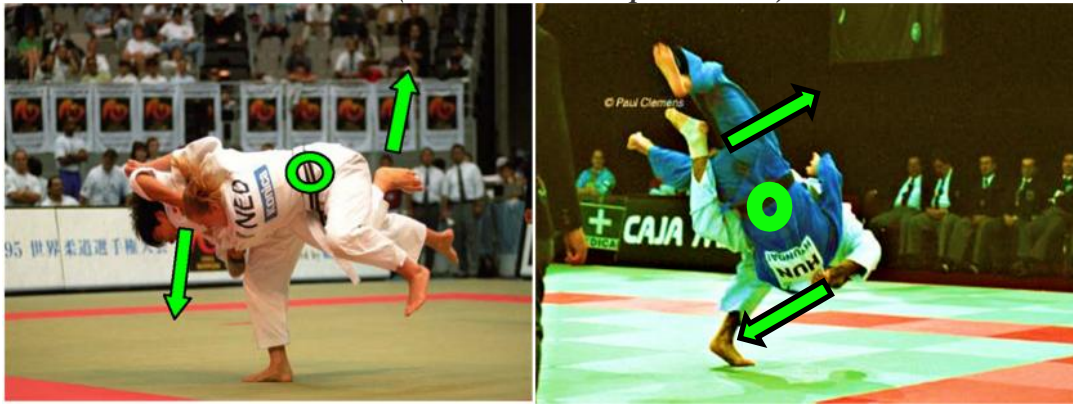

*Fig.4,5 annexes Uchi-mata and Harai goshi are typical technique of Couple forces applied by trunk and leg [8] (David Finch with permission).*

| | | | |
|---|---|---|---|
| **"Couple of Forces"-type Throwing Techniques**<br><br>Couple applied by: | Arms | Kuchiki-daoshi<br>Kibisu-gaeshi<br>Kakato-gaeshi<br>Te-guruma   Uchi mata sukashi | **All Innovative Variations of Throws and very few Chaotic Forms of Throws** |
| | Arm/s and leg | De-ashi-barai, -uchi-gari, Okuri-ashi-barai, Ko-uchi-gake, Ko-uchi-barai, Ko-soto-gake, -uchi-barai, Harai-tsuri-komi-ashi, Tsubame-gaeshi, Yoko-gake, Ko-uchi-gari, -soto-gake, Ko-soto-gari, -uchi-gake, O-uchi-gaeshi(1) | |
| | Trunk and legs | -soto-gari, -tsubushi, -soto-guruma, -soto-otoshi, Uchi-mata, Ko-uchi-sutemi, Okuri-komi-uchi-mata, Harai-makikomi, Harai-goshi, Ushiro-uchi-mata, Ushiro-hiza-ura-nage, Hane-goshi, Gyaku-uchi-mata, Hane-makikomi, Daki-ko-soto-gake, Yama-arashi (Khabarelli-type throw), Uchi-Mata-gaeshi, Hane-goshi gaeshi, Harai-Goshi -gaeshi,Uchi-Mata-makikomi, Harai-makikomi, Hane-makikomi, | |
| | Trunk and arms | Morote-gari | |
| | Legs | Kani-basami | |

*Tab 1 Annexes   Couple Techniques    The table is based on the most recent Kodokan classification [11] and innovative techniques names [12]*

| Physical Lever-type Throwing Techniques Lever applied by: | **Minimum Arm Lever** (fulcrum under *uke*'s waist) | -guruma, Ura-nage, Kata-guruma, Ganseki-otoshi, Tama-guruma, Uchi-makikomi, Binta Guruma, Obi-otoshi, Soto-Makikomi, Tawara-gaeshi, Makikomi, Kata-sode-ashi-tsuri, Sukui-nage, Daki-sutemi, Ushiro-goshi, Utsuri-goshi | **All Innovative Variation and New (Chaotic) Forms** |
|---|---|---|---|
| | **Medium Arm Lever** (fulcrum under *uke*'s knees) | *Hiza-guruma, Ashi-guruma, Hiza-soto-mus , Soto-kibisu-gaeshi* | |
| | **Maximum-Arm Lever** (fulcrum under *uke*'s malleolus) | *Uki-otoshi, Yoko-guruma, Yoko-otoshi, Yoko-wakare, Sumi-otoshi, Seoi-otoshi, Suwari-otoshi, Hiza-seoi, No -Waki, O-uchi-gaeshi(2) Waki-otoshi, Obi-seoi, Tani-otoshi, Suso-seoi, Tai-otoshi, Suwari-Seoi, Dai-sharin, Hiza-tai-otoshi, Hikkomi-gaeshi, Tomoe-nage, Sumi-gaeshi, Ry -ashi-tomoe, Yoko-kata-guruma, Yoko-tomoe, Uki-waza, Sasae-tsuri-komi-ashi, Uke-nage* | |
| | **Variable Arm** (variable fulcrum from *uke*'s waist to his knees) | *Tsuri-komi-goshi, Kubi-nage, -goshi, Sasae-tsuri-komi-goshi, Koshi-guruma, Ko-tsuri-komi-goshi, -tsuri-komi-goshi, Sode-tsuri-komi-goshi Seoi-nage, Eri-seoi-nage, Uki-goshi, Morote-seoi-nage* | |

*Tab 2 Annexes Lever Techniques The table is based on the most recent Kodokan classification [11] and innovative techniques names [12]*